# Real-Time Dynamics of Triplet-Resonant Tunneling Driven by Nonequilibrium Phonons

Kazuyuki Kuroyama[1,*], Sasha R. Valentin[2], Arne Ludwig[2], Andreas D. Wieck[2], Yasuhiro Tokura[3], Seigo Tarucha[4,*], and Sadashige Matsuo[4,5,*]

[1] Institute of Industrial Science, The University of Tokyo, Tokyo, Japan

[2] Lehrstuhl für Angewandte Festkörperphysik, Ruhr-Universität Bochum, D-44780 Bochum, Germany

[3] Pure and Applied Sciences, University of Tsukuba, 1-1-1 Tennodai, Tsukuba, Ibaraki 305-8571, Japan

[4] Center for Emergent Materials Science (CEMS), RIKEN, 2-1 Hirosawa, Wako-shi, Saitama 351-0198, Japan

[5] Department of Physics, Institute of Science Tokyo, 2-12-1 Ookayama, Tokyo, Japan

*corresponding author: kuroyama@iis.u-tokyo.ac.jp, tarucha@riken.jp, matsuo@phys.sci.isct.ac.jp

## Abstract

Driven nonequilibrium systems can host emergent functionalities beyond equilibrium, but real-time access to excited-state dynamics remains limited. Here we report real-time measurements of phonon-driven charge and spin dynamics in excited states of a double quantum dot. Under phonon irradiation, resonant inter-dot tunneling emerges at triplet resonance. Time-resolved charge sensing reveals that the resonant inter-dot tunneling is strongly modified by spin blockade. For weaker inter-dot coupling, the nonequilibrium phonon environment generates a unidirectional transport cycle along the phonon density gradient.

Nonequilibrium phenomena in solids provide a powerful route to functionalities and emergent responses that have no equilibrium counterpart [1,2]. In driven systems, a continuous flow of energy and particles, balanced by dissipation into the environment, can stabilize nonequilibrium steady states and generate responses qualitatively distinct from those in equilibrium. Correlated electron systems under nonequilibrium conditions have attracted considerable attention as a prominent platform for such physics. The Hubbard model [3] has long served as a paradigmatic framework for describing interaction-driven phases such as Mott insulators, magnetism, and superconductivity [4]. While much of this effort has focused on equilibrium properties, growing attention has turned to driven Hubbard systems subjected to optical or phononic excitation, where nonequilibrium phase transitions and emergent states beyond equilibrium have been theoretically proposed [5–7]. In such nonequilibrium Hubbard-type systems, an essential issue is not only the behavior of the many-body ground state but also the dynamics involving excited states. In particular, to experimentally clarify nonequilibrium phase transitions in Hubbard-type systems, the ability to probe excited-state dynamics in real time and with local resolution would provide spatiotemporal access to the properties of condensed phases in nonequilibrium systems.

Semiconductor quantum dots (QDs) [8] offer an exceptionally well-controlled mesoscopic platform for investigating Hubbard-type models at the single-charge level [9–11]. Their high tunability and scalability also make them a promising platform for quantum simulation of correlated electron systems. In particular, recent advances in time-resolved charge detection now enable real-time access to single-electron dynamics in nanostructures [12,13]. Furthermore, recent studies have shown that photon [14,15] or phonon [16–18] irradiation can promote electrons to higher-energy orbital states in QDs. By combining phonon-driven excitation with time-resolved and spatially local charge sensing, QD platforms provide direct access to charge and spin dynamics in excited states under a driven nonequilibrium environment.

Here we report real-time observations of resonant excited-state dynamics in a semiconductor double quantum dot (DQD). Under phonon irradiation of the DQD, we observe interdot resonant tunneling at the triplet resonance, corresponding to a resonance between excited states. Using a charge sensor, we perform time-resolved measurements of DQD charge-state transitions at the triplet resonance. We investigate the behavior of this triplet-mediated interdot resonant tunneling in regimes where the interdot tunneling rate is larger than or comparable to the dot–reservoir tunneling rates. In the former regime, we observe repeated interdot tunneling events mediated by the triplet-state resonance, which are strongly modified by spin blockade. Furthermore, in the latter, weak-coupling regime, we find that a nonequilibrium phonon environment generates a unidirectional electron-transport cycle along the phonon density gradient. Our real-time measurements of a phonon-driven DQD system provide a microscopic basis for understanding and controlling driven nonequilibrium steady states in mesoscopic systems.

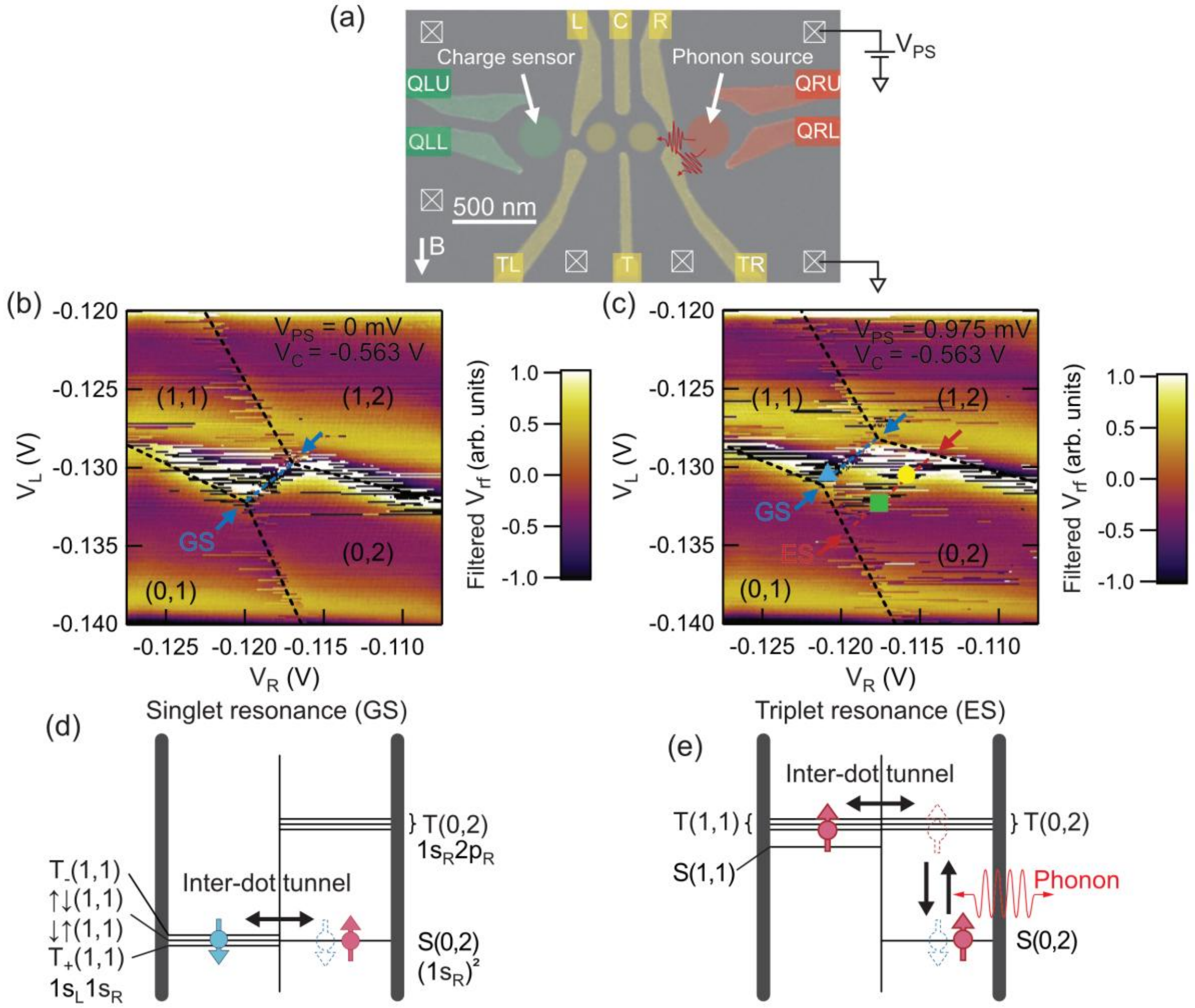

FIG. 1. (a) Geometry of our DQD sample with a charge sensor and a QD-based phonon source. (b) Charge stability diagram of the DQD near the (1,1)-(0,2) inter-dot resonance at $V_{\mathrm{PS}} = 0$ mV. The singlet resonance line is guided by the blue dashed line, and the boundaries of the charge states are guided by the black dashed lines. (c) Charge stability diagram measured at $V_{\mathrm{PS}} = 0.975$ mV. The singlet (GS) and triplet (ES) resonance lines are indicated by the blue and red dashed lines. (d,e) Transition diagrams at the singlet (d) and triplet (e) resonance conditions. $1\mathrm{s_L}$ and $1\mathrm{s_R}$ denote the 1s orbitals in the left and right QDs, respectively, and $2\mathrm{p_R}$ denotes the 2p orbital in the right QD.

Our DQD sample was fabricated on a modulation-doped AlGaAs/GaAs/AlGaAs double heterostructure wafer. Two-dimensional electrons are accumulated in the 15-nm-thick GaAs quantum well located 105 nm below the wafer surface. The geometry of the surface gate electrodes is shown in Fig. 1(a). The DQD is formed in the quantum well layer by the split gates colored yellow. The DQD charge state is monitored in real time by a nearby charge sensor QD located on the left side of the DQD. The electrical conductance of the charge sensor was measured using a radio-frequency resonator circuit by detecting the voltage amplitude $V_{\mathrm{rf}}$ of the radio-frequency wave reflected from the circuit. On the opposite side of the DQD, we formed another QD as indicated by the red circle in Fig. 1(a), which works as a phonon source [18,19]. We applied a DC source-drain bias voltage $V_{\mathrm{PS}}$ to this phonon source to generate acoustic phonons. Note that no DC bias voltage is applied to the sensor QD so that phonons are generated only from the phonon source QD. The DQD sample was measured in a dilution refrigerator with a base temperature of about 20 mK. All the experiments described in the following part of this paper were performed under an in-plane magnetic field of 100 mT, which was applied parallel to the quantum well layer [see Fig. 1 (a) for the magnetic field direction].

To characterize the DQD, we measured the charge stability diagrams near the inter-dot resonance of the (1,1) and (0,2) charge states. Figure 1(b) shows the charge sensing signal $V_{\rm rf}$ measured as a function of the gate voltages $V_{\rm L}$ and $V_{\rm R}$ applied to the L and R gates, respectively, without a DC bias voltage applied to the phonon source QD, i.e., $V_{\rm PS} = 0$ mV. No source-bias voltage is applied across the DQD. Note that the background oscillation originated from the Coulomb oscillation of the sensor QD is removed by applying a low-pass filter. The boundaries between the different charge states are guided by blue and black dashed lines in the figure. The blue dashed line marks the inter-dot resonance condition between the (1,1) charge state and the ground state of the (0,2) charge state. The energy diagram of the electron spin states of the DQD in the inter-dot resonance condition is pictorially explained in Fig. 1(d). In the two-electron DQD, the energy ground state of the (0,2) charge state is the spin-singlet $|S(0,2)\rangle$ and the ground states of the (1,1) charge state consists of four different spin states that are nearly degenerate compared to the thermal energy. The two spin states with anti-parallel spins are denoted by $|\uparrow\downarrow(1,1)\rangle$, and $|\downarrow\uparrow(1,1)\rangle$, and the other spin states with parallel spins are denoted by $|T_{+}(1,1)\rangle$, and $|T_{-}(1,1)\rangle$, where $|T_{\nu}(1,1)\rangle$ $(\nu = \pm)$ are spin triplet states. Note that $|S(1,1)\rangle$ and $|T_0(1,1)\rangle$ are not the eigenstates because of the weak inter-dot tunnel coupling and the hyperfine interaction with nuclear fields [20]. At the inter-dot resonant condition, an electron can resonantly tunnel between $|S(0,2)\rangle$ and $|\uparrow\downarrow(1,1)\rangle$ or $|\downarrow\uparrow(1,1)\rangle$.

Next, we turned on the phonon source QD by applying $V_{\rm PS}$. The biased phonon-source QD emits nonequilibrium acoustic phonons locally into the host crystal. Because the source is placed on the right side of the DQD, the resulting phonon flux creates a spatial phonon-density gradient across the two dots [18,19]. Figure 1(c) shows the charge-stability diagram taken in the same gate-voltage region at $V_{\rm PS} = 0.975$mV. Besides the ground-state singlet resonance [blue dashed line in Fig. 1(b)], an additional resonance [red dashed line in Fig. 1(c)] emerges in the (0,2) region. This feature corresponds to a triplet resonance between the $|T(1,1)\rangle$ manifold and the first orbital-excited $|T(0,2)\rangle$, as illustrated in Fig. 1(e). Phonon absorption promotes an electron in the right dot from the 1s to the 2p orbital [15,19,21,22]; in the presence of spin–orbit coupling, this excitation accesses the $|T(0,2)\rangle$ manifold. When the electrochemical potentials of the two triplet configurations align, resonant inter-dot tunneling repeatedly occurs between the corresponding $|T(1,1)\rangle$ and $|T(0,2)\rangle$ states, giving rise to the second resonance line.

We performed time-resolved measurements of the DQD charge state at the singlet and triplet resonance conditions [see Figs. 1(d) and 1(e)]. We note that, by applying a voltage to gate C, the interdot tunneling rate was made larger than the dot–reservoir tunneling rates. In this regime, the DQD can be regarded as an effectively isolated system. As a reference, Fig. 2(a) shows a typical time trace measured at the singlet resonance (GS) under phonon irradiation with $V_{\rm PS} =$ 1.3 mV. The fast charge state transitions between (1,1) and (0,2) are attributed to resonant inter-dot tunneling between $|\uparrow\downarrow(1,1)\rangle$ and $|S(0,2)\rangle$ and between $|\downarrow\uparrow(1,1)\rangle$ and $|S(0,2)\rangle$ as shown in Fig. 1(d). Occasionally, the system dwells in (1,1) for a relatively long time. This behavior can be explained by the Pauli spin-blockade of the parallel spin states in the DQD. $|T_{+}(1,1)\rangle$ and $|T_{-}(1,1)\rangle$ cannot tunnel to (0,2) due to this effect [18,20,23–25]. A subsequent spin-flip process lifts the blockade and restores the fast (1,1)-(0,2) inter-dot switching.

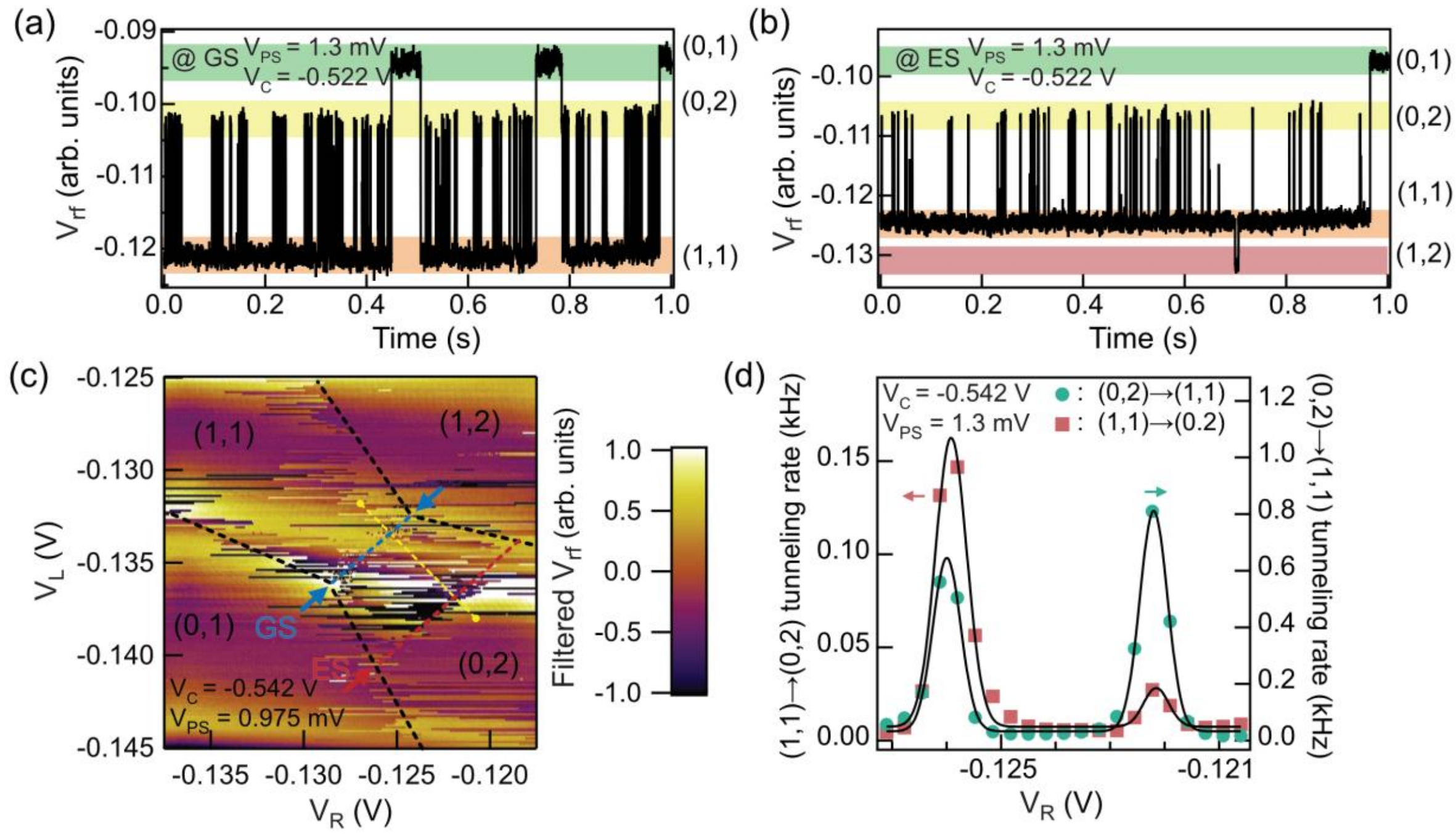

FIG. 2. (a) Typical time trace of inter-dot tunneling events measured at the singlet resonance. (b) Typical time trace of inter-dot tunneling events measured at the triplet resonance. (c) Charge stability diagram measured under phonon irradiation at slightly different inter-dot tunnel coupling strength from Fig. 1(c) by changing a gate voltage applied to gate C. (d) Inter-dot tunnel rates of the (1,1)→(0,2) tunneling (red squares) and the (0,2)→(1,1) tunneling (green circles) extracted along the yellow dashed cut in (c). The left (right) peak corresponds to the singlet (triplet) resonance.

Next, we tuned the DQD operation point to the triplet resonance (ES) without changing the inter-dot tunneling barrier and measured a time trace under the same phonon irradiation with $V_{\mathrm{PS}} = 1.3$ mV as shown in Fig. 2(b). While the inter-dot tunneling is still observed, the system is more likely to remain in (1,1) compared with the singlet-resonance case.

To verify that the second resonance (red dashed line) indeed reflects resonant inter-dot tunneling, we measured the inter-dot transition rates as a function of detuning along the cut indicated by the yellow dashed line in Fig. 2(c). Note that the charge stability diagram was measured at slightly different gate voltage applied to the C gate from Fig. 1(c). Figure 2(d) summarizes the rates extracted from 500-s time traces. We define the inter-dot tunnel rate of (1,1)→(0,2) [(0,2)→(1,1)] as the number of identified (1,1)→(0,2) [(0,2)→(1,1)] events divided by the total dwell time in (1,1) [(0,2)]. In Fig. 2(d), the (1,1)→(0,2) rate (red squares) and the (0,2)→(1,1) rate (green circles) are plotted as a function of $V_{\mathrm{R}}$. Both rates exhibit two pronounced peaks. The peak at more negative $V_{\mathrm{R}}$ corresponds to the singlet resonance, while the peak at less negative $V_{\mathrm{R}}$ corresponds to the triplet resonance. The black solid line is a fit to the sum of two Gaussian functions. The fact that both directional rates peak at nearly the same detuning confirms that these features arise from resonant inter-dot tunneling.

To investigate the inter-dot tunneling at the triplet resonance, we analyze the waiting-time statistics of inter-dot tunneling events. From 500-s time traces, we construct histograms of the dwell time in a given charge state before an inter-dot transition occurs, separately for (1,1) and (0,2). At the singlet-resonance condition, the waiting-time histograms for (0,2) and (1,1) are shown in Figs. 3(a) and 3(b), respectively. The (0,2) histogram in Fig. 3(a) shows a single exponential distribution, consistent with previous reports [18,20,23–25]. We fit the distribution with a single exponential as shown by the red curve in Fig. 3(a), and the estimated transition rate is $\Gamma_{02}^{GS} = 1.76 \pm 0.02$ kHz. The (0,2) rate $\Gamma_{02}^{GS}$ reflects resonant tunneling from $|S(0,2)\rangle$ into the (1,1) manifold. In contrast, the (1,1) histogram of Fig. 3(b) exhibits a clear double exponential behavior, also consistent with previous reports. We fit the distribution with a sum of two exponentials as shown by the red curve in Fig. 3(b). The transition rates of the faster and slower transitions estimated from the fitting curve are $\Gamma_{11,f}^{GS} = 890.9 \pm 4.3$ Hz and $\Gamma_{11,s}^{GS} = 68.6 \pm 9.1$ Hz. We interpret $\Gamma_{11,f}^{GS}$ as the rate of spin-conserving resonant tunneling and $\Gamma_{11,s}^{GS}$ as the spin-flip–assisted process that lifts Pauli blockade [18,20,23–25].

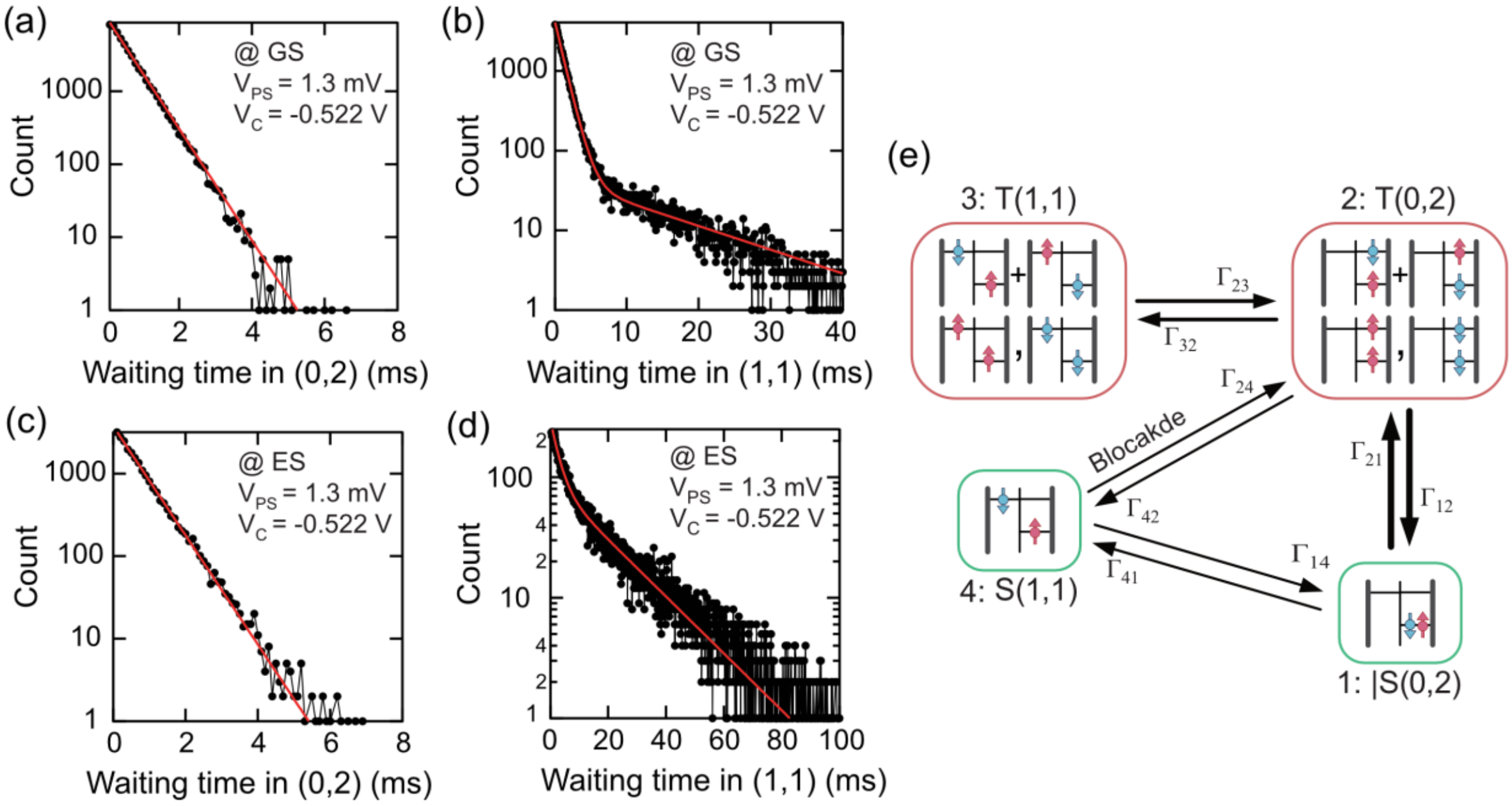

FIG. 3. (a, b) (0,2) and (1,1) waiting-time histograms obtained from the time traces measured at the singlet-resonance (GS) are plotted, respectively. The red curves are fits to a single-exponential function in (a) and a double-exponential function in (b). (c,d) (0,2) and (1,1) waiting-time histograms obtained from the time traces measured at the triplet resonance condition (ES). The red curves are fits to a single-exponential function in (c) and a double-exponential function in (d). (e) DQD transition diagram at the triplet resonance. The spin state initially located at $|S(0,2)\rangle$ is frequently excited to $|T(0,2)\rangle$ by a single phonon excitation. When the DQD spin state enters $|S(1,1)\rangle$, the system is trapped in the (1,1) charge state due to the Pauli blockade.

We performed the same waiting-time analysis for time traces of the DQD charge state transitions observed at the triplet-resonance condition. The waiting-time histogram for (0,2) and (1,1) are shown in Fig. 3(c) and 3(d), respectively. Similar to the singlet-resonance case, the (1,1) histogram shows a clear double-exponential distribution, while the (0,2) histogram shows a clear single-exponential distribution. The red curve in Fig. 3(c) is the fitted single-exponential function, and the estimated transition rate is $\Gamma_{02}^{ES} = 1.52 \pm 0.01$ kHz. The fact of $\Gamma_{02}^{ES} \sim \Gamma_{02}^{GS}$ implies that $\Gamma_{02}^{ES}$ is dominantly determined by the resonant tunneling at the triplet resonance. The red curve in Fig. 3(d) is the fitted double-exponential function. The transition rates of the faster and slower transitions estimated from the fitting curve are $\Gamma_{11,f}^{ES} = 361.9 \pm 12.2$ Hz and $\Gamma_{11,s}^{ES} = 54.4 \pm 1.7$ Hz, respectively.

To explain the physical mechanism behind the transition rates $\Gamma_{11,\mathrm{f}}^{\mathrm{ES}}$ and $\Gamma_{11,\mathrm{s}}^{\mathrm{ES}}$, we consider the transition diagram shown in Fig. 3(e). At the triplet-resonance condition, after phonon absorption excites $|S(0,2)\rangle$ to $|T(0,2)\rangle$, resonant spin-conserving inter-dot tunneling takes place between $|T(1,1)\rangle$ and $|T(0,2)\rangle$. Moreover, the tunneling process between $|S(1,1)\rangle$ and $|T(0,2)\rangle$ is also possible by the spin-flip tunneling [20]; however, because of the orthogonality of these spin states, this tunneling process is suppressed and less frequent due to the Pauli spin blockade. This blockade manifests as events with long dwell times in the (1,1) charge state in the time trace shown in Fig. 2(b). Therefore, the faster and slower inter-dot component are attributed to the resonant tunneling between $|T(1,1)\rangle$ and $|T(0,2)\rangle$ and to Pauli spin blockade of $|S(1,1)\rangle$, respectively.

According to the spin transition rates $\Gamma_{ij}$ defined in the transition diagram of Fig. 3(e), we derive analytical expressions for the waiting-time histograms of the (1,1) and (0,2) charge states. We introduce the experimental time resolution $\delta t$ of our time-trace measurements and the total dwell time $T_{\mathrm{tot}}$ in the (1,1) or (0,2) charge state. In addition, we assume that the phonon-assisted transitions between $|S(0,2)\rangle$ and $|T(0,2)\rangle$ are much faster than all other transitions. Under this approximation, the waiting-time histograms are written as

$$W_{02}(T) = \frac{\delta t T_{\mathrm{tot}} \Gamma_{02}}{D}[w(\Gamma_{32} + \Gamma_{41}) + (1-w)\Gamma_{42}]e^{-\Gamma_{02}T}, \quad (1)$$

$$W_{11}(T) = \frac{\delta t T_{\mathrm{tot}}}{D}\big[(1-w)\Gamma_{23}\Gamma_{32}e^{-\Gamma_{23}T} + (\Gamma_{14} + \Gamma_{24})[w\Gamma_{41} + (1-w)\Gamma_{42}]e^{-(\Gamma_{14}+\Gamma_{24})T}\big], \quad (2)$$

where

$$\Gamma_{02} = w\Gamma_{41} + (1-w)(\Gamma_{32} + \Gamma_{42}), \quad (3)$$

$$D = 1 + \frac{w\Gamma_{32}}{\Gamma_{23}} + \frac{w\Gamma_{41} + (1-w)\Gamma_{42}}{\Gamma_{14} + \Gamma_{24}}. \quad (4)$$

Here, $w = \Gamma_{12}/(\Gamma_{12} + \Gamma_{21})$ and $1-w$ denote the population fraction in $|S(0,2)\rangle$ and in $|T(0,2)\rangle$, respectively. A detailed derivation of the waiting-time distributions is provided in the Supplemental Material [url inserted later]. These expressions show that the waiting-time distribution in the (0,2) charge state is single exponential, whereas that in (1,1) is a sum of two exponentials, consistent with the measured histograms in Figs. 3(c) and 3(d). We note, however, that Eqs. (1) and (2) do not uniquely determine all transition rates, because the number of independent fit parameters exceeds the information contained in the histograms. Moreover, as shown in Fig. 2(b), because the blockade dwell time is longer than the characteristic time of the spin-conserving transition, we infer that $\Gamma_{42}$ is larger than either $\Gamma_{24}$ or $\Gamma_{14}$. This asymmetry between the spin-flip inter-dot tunnel rates of $\Gamma_{42}$ and $\Gamma_{24}$ may originate from the nonequilibrium phonon environment in the DQD.

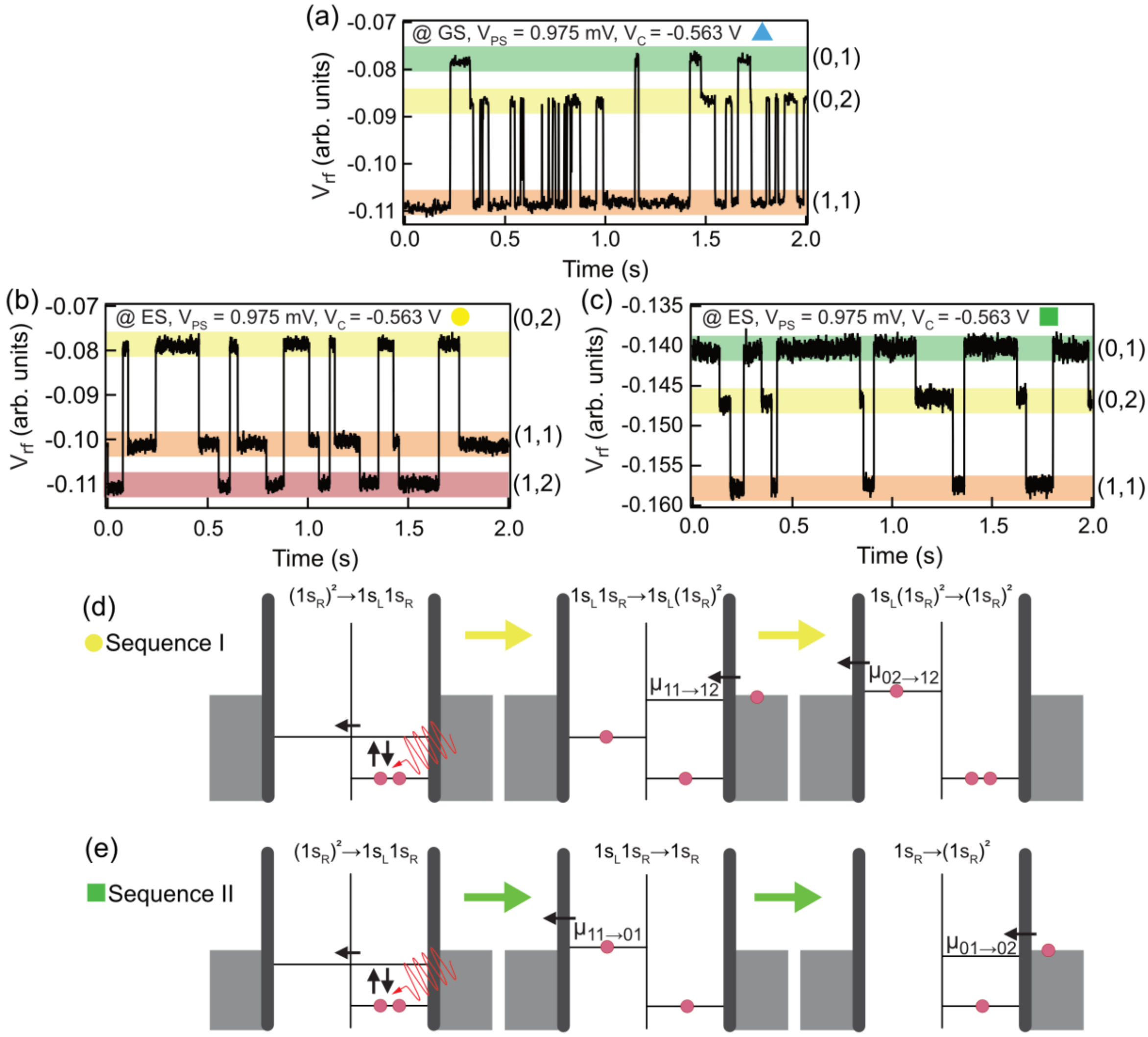

FIG. 4. (a) Typical time trace of the DQD inter-dot tunneling events observed at the singlet resonance for reduced inter-dot tunnel coupling, tuned by applying a more negative voltage to gate C (see Fig. 1(a) for the sample geometry). (b, c) Typical time trace of inter-dot tunneling events measured at the triplet resonance marked by the yellow circle and the green square in Fig. 1(c). The time trace in (b) is measured at the DQD operation point closer to the (1,2) region, while the time trace in (c) is measured at the DQD operation point closer to the (0,1) region. (d,e) Schematic sequences illustrating phonon-driven unidirectional charge-transfer cycles: (d) (0,2)→(1,1)→(1,2)→(0,2) and (e) (0,2)→(1,1)→(0,1)→(0,2). The cycles correspond to the operating points marked by the yellow circle (d) and green square (e) in Fig. 1(c).

It is worth noting that $|S(1,1)\rangle$ and $|T_0(1,1)\rangle$ can be readily mixed by the difference in the nuclear fields between the two dots [26,27], particularly when the inter-dot tunnel coupling is small. In contrast, the nuclear fields in the two dots can be nearly identical due to several mechanisms [28], e.g., nuclear-spin pumping [29]. In our case, the origin of the suppressed $|S(1,1)\rangle$ -$|T_0(1,1)\rangle$ mixing remains unclear, but it may be related to phonon-assisted nuclear-spin polarization.

Finally, we discuss unidirectional electron tunneling in the triplet states under phonon irradiation at weaker inter-dot coupling. We reduce the inter-dot tunnel coupling by applying a more negative voltage to gate C, $V_{\mathrm{C}} = -0.563$ V, where the DQD inter-dot tunnel coupling strength is comparable to the dot-reservoir coupling. It should be noted that no source-bias voltage is applied across the DQD. As a reference, we measured a time trace of the DQD charge states at the singlet resonance condition under phonon irradiation as shown in Fig. 4(a). Except for the decrease in the tunnel rates, the inter-dot dynamics at the singlet resonance are essentially the same as in Fig. 2(a).

We then tune the DQD operating point to the triplet resonance condition, closer to the (1,2) region, as indicated by the yellow circle in Fig. 1(c). The corresponding time trace in Fig. 4(b) does not show the repeated inter-dot tunneling; instead, the system undergoes a unidirectional charge state transition (1,1)→(1,2)→(0,2)→(1,1). Figure 4(c) also shows a time trace taken at the triplet resonance condition closer to the (0,1) region. We observe a unidirectional transition (0,1)→(0,2)→(1,1)→(0,1). In both cases, each cycle transfers one electron from the right to the left reservoir along the phonon-density gradient, even in the absence of a source–drain bias across the DQD.

To quantify the rectification, we define the transfer asymmetry

$$\eta = \frac{|N_{\mathrm{R}\to\mathrm{L}} - N_{\mathrm{L}\to\mathrm{R}}|}{N_{\mathrm{R}\to\mathrm{L}} + N_{\mathrm{L}\to\mathrm{R}}}, \tag{5}$$

where $N_{\mathrm{R}\to\mathrm{L}}$ ($N_{\mathrm{L}\to\mathrm{R}}$) is the number of single-electron transfer events from the right (left) to the left (right) reservoir. From 400-s traces, we obtain $\eta = 0.939$ ($N_{\mathrm{R}\to\mathrm{L}} = 825, N_{\mathrm{L}\to\mathrm{R}} = 26$) for Fig. 4(b) and $\eta = 0.997$ ( $N_{\mathrm{R}\to\mathrm{L}} = 825, N_{\mathrm{L}\to\mathrm{R}} = 1$ ) for Fig. 4(c), demonstrating strong rectification of the phonon-driven charge transfer. The current $I$ induced by the nonequilibrium phonon environment without the DQD source-drain bias voltage is also evaluated as $3.2 \times 10^{-19}$ A for Fig. 4(b) and $3.3 \times 10^{-19}$ A for Fig. 4(c), corresponding to about 2 electrons per second.

The observed unidirectional tunneling is enabled when nonequilibrium phonon irradiation excites the DQD from (0,2) to (1,1). Figure 4(d) explains the sequential tunneling processes of the unidirectional cycle (0,2)→(1,1)→(1,2)→(0,2) (Sequence I). The cycle starts when phonon absorption excites $|S(0,2)\rangle$ to $|T(0,2)\rangle$, and resonant tunneling to $|T(1,1)\rangle$. The system is likely to remain trapped in (1,1), implying that the return rates $\Gamma_{14}$ and $\Gamma_{24}$ are small compared with $\Gamma_{41}$ and $\Gamma_{42}$. This suppression is attributed to the reduced phonon flux reaching the left dot and to Pauli spin blockade associated with the $|S(1,1)\rangle$ configuration. When the system is in the (1,1) charge state, the electrochemical potential for the (1,1)→(1,2) transition, $\mu_{11\to12}$, becomes comparable with the Fermi level of the right reservoir, allowing an electron to enter the right QD. Subsequently, an electron in the left QD can tunnel out to the left reservoir because the electrochemical potential for the (0,2)→(1,2) transition, $\mu_{02\to12}$, lies above the Fermi energy of the left reservoir. An electron tunnels out from the left dot to the left reservoir, returning the DQD to (0,2). Overall, one electron is transferred from the right to the left reservoir per cycle. An analogous argument applies to sequence II, (0,2)→(0,1)→(1,1)→(0,2), at the operating point marked by the green square in Fig. 1(c), as summarized in Fig. 4(e). Our real-time measurements reveal nonreciprocal transport emerging at a resonance under nonequilibrium phonon driving. This establishes an experimentally accessible route to microscopic studies of nonequilibrium nonreciprocity and rectification in quantum-dot platforms.

We observe inter-dot electron tunneling mediated by triplet states in a double quantum dot (DQD) driven by nonequilibrium phonon irradiation. Using time-resolved charge sensing, we observe tunneling dynamics at the triplet-resonance condition in the time domain. Waiting-time statistics extracted from long time traces are captured by a rate-equation model for the DQD spin states, yielding single- and double-exponential waiting-time distributions for (0,2) and (1,1), respectively. Furthermore, at reduced inter-dot tunnel coupling, nonequilibrium phonon irradiation induces strongly rectified charge-transfer cycles at the triplet resonance, resulting in unidirectional single-electron transfer even without an applied source–drain bias across the DQD. These results provide a microscopic time-domain view of nonequilibrium phonon-driven transport and suggest routes toward nonequilibrium phonon-driven quantum functionalities.

## Acknowledgement

This work was supported by KAKENHI from JSPS (JP22K03481), JST, PRESTO (JPMJPR2255), JST FOREST (JPMJFR223A), the Murata Science and Education Foundation, the CANON foundation, the Asahi Glass foundation, and the Fujikura Foundation.

## Data availability

The data that support the findings of this article are not publicly available. The data are available from the authors upon reasonable request.